\documentclass{kapproc} 
\usepackage{procps}
\usepackage[dvips]{graphicx}
\upperandlowercase
\kluwerbib

\def\simlt{\mathrel{\rlap{\lower 3pt\hbox{$\sim$}} \raise
        2.0pt\hbox{$<$}}} \def\simgt{\mathrel{\rlap{\lower
        3pt\hbox{$\sim$}} \raise 2.0pt\hbox{$>$}}}

\begin{document}

\articletitle[The Halo Occupation Number and Spatial Distribution of 2dF 
Galaxies]
{The Halo Occupation Number and Spatial Distribution of 2dF Galaxies}

\author{Manuela Magliocchetti$^1$ \& Cristiano Porciani$^2$}
\affil{$^1$SISSA, Via Beirut 4, 34014, Trieste, Italy\\
$^2$Institute of Astronomy, HPF G3.2, ETH Hoenggerberg, 8093 Zuerich,
Switzerland}

\anxx{Magliocchetti\, Manuela}

\begin{abstract}
We use the clustering results obtained by Madgwick et al. (2003) for a sample 
of 96,791 2dF galaxies with redshift $0.01 < z < 0.15$ to study the 
distribution of late-type and early-type galaxies within dark matter haloes 
of different mass. The adopted method relies on the connection between the 
distribution of sources within haloes and their clustering properties 
by focussing on the issue of the halo occupation function i.e. 
the probability distribution of the 
number of galaxies brighter than some luminosity threshold hosted 
by a virialized halo of given mass. 
Within this framework, the distribution of galaxies within haloes 
is shown to determine galaxy-galaxy clustering on small scales, being 
responsible for the observed power-law behaviour at separations
$r\simlt 3$~Mpc. For a more extended analysis, we refer the reader 
to Magliocchetti \& Porciani (2003).
\end{abstract}


\section{Analysis and Results}

Our approach follows the one adopted by Scoccimarro et al. (2001);
in this framework, the galaxy-galaxy correlation function can be written as \\
\begin{eqnarray}
\xi_g({\bf x}-{\bf x}^\prime)=\xi_g^{1h}({\bf x}-
{\bf x}^\prime)+\xi_g^{2h}({\bf x}-{\bf x}^\prime),
\label{eq:xi}
\end{eqnarray}
where the first term $\xi_g^{1h}$ accounts for pairs of galaxies residing 
within 
the same halo, while the $\xi_g^{2h}$ represents the contribution coming 
from galaxies in different haloes. The above quantities can be written as 
a function of the mean number of galaxies per halo of mass $m$, 
$\langle N_{\rm gal}(m) \rangle$, of the spread about 
this mean value (also dependent on the mass of the halo hosting the galaxies), 
$\langle N^2_{\rm gal}(m)\rangle$, of the mean comoving number density of galaxies, 
$\bar{n}_g=\int n(m) \langle N_{\rm gal}(m) \rangle dm$ (where $n(m)$ is the 
halo mass function which gives the number density of dark matter 
haloes per unit mass and volume), of 
the two-point cross-correlation function between haloes of mass 
$m_1$ and $m_2$, $\xi(r, m_1, m_2)$ and  
finally of the (spatial) density distribution of galaxies 
within the haloes, $\rho_m(r)$.

$\langle N_{\rm gal}(m) \rangle$ and $\langle N^2_{\rm gal}(m)\rangle$ are 
the first and second moment of the halo occupation function $p(N_{\rm gal}|m)$ 
which gives the probability 
for a halo of specified mass $m$ to contain $N_{\rm gal}$ galaxies. These 
can be parameterized as  $\langle N_{\rm gal}\rangle=0,(m/m_0)^{\alpha_1},
(m/m_0)^{\alpha_2}$ and $\langle N^2_{\rm gal}\rangle=\alpha(m)^2 
(\langle N_{\rm gal}\rangle)^2$ with $\alpha(m)=0,\\
{\rm log}(m/m_{\rm cut})
/{\rm log}(m_0/m_{\rm cut}),1$ in the different mass ranges 
$m< m_{\rm cut}$, $m_{\rm cut}\le m < m_0$ and $m\ge m_0$.
According to this approach, $m_{\rm cut}$, $m_0$, $\alpha_1$ and $\alpha_2$ 
are parameters to be determined by comparison with observations.

The last ingredient needed for 
the description of 2-point galaxy clustering is the spatial distribution
of galaxies within their haloes. The first, easiest approach one can take is to assume 
that galaxies follow the dark matter profile ( Navarro, Frenk \& White 1997, 
hereafter NFW). However, since this assumption is not necessarily true,
in the present analysis we also consider spatial 
distributions of the form $\rho_m(r)\propto (r)^{-\beta}$,
with $\beta=2,2.5,3$, where the first value corresponds to the singular 
isothermal sphere case. All the above profiles have been initially truncated 
at the virial radius $r_{\rm vir}$, as one expects galaxies to form within 
virialized regions, where the overdensity is greater than a certain threshold. 
However, this might not be the only possible choice since for instance -- 
as a consequence of halo-halo merging -- galaxies might also be found in the outer 
regions of the newly-formed halo, at a distance from the center greater than 
$r_{\rm vir}$.

In order to perform our analysis, we have considered the 4-dimensional grid:
$-1 \le \alpha_1\le 2$; $-1 \le \alpha_2\le 2$; $10^9\; m_\odot\le m_{{\rm cut}} 
\le 10^{13}$; $m_{{\rm cut}}\le m_0\le m_{{\rm cut}}\cdot 10^3 $.
Combinations of these four quantities have then been used to evaluate 
the mean number density of galaxies $\bar{n}_g$. Only values for $\bar{n}_g$ 
within 2$\sigma$ from the observed ones (derived from the luminosity functions
obtained by Madgwick et al. 2002) were accepted and the corresponding values for 
$\alpha_1$, $\alpha_2$, $m_{\rm cut}$ and $m_0$ have subsequently been 
plugged into equation (\ref{eq:xi}) to produce -- for a specified choice of 
the distribution profile -- the predicted 
galaxy-galaxy correlation function to be compared with the Madwick et al. (2003) 
results on late-type and early-type galaxies by means of a least squares ($\chi^2$) 
fit.

\begin{figure}[ht]
\vskip2.7in
\includegraphics{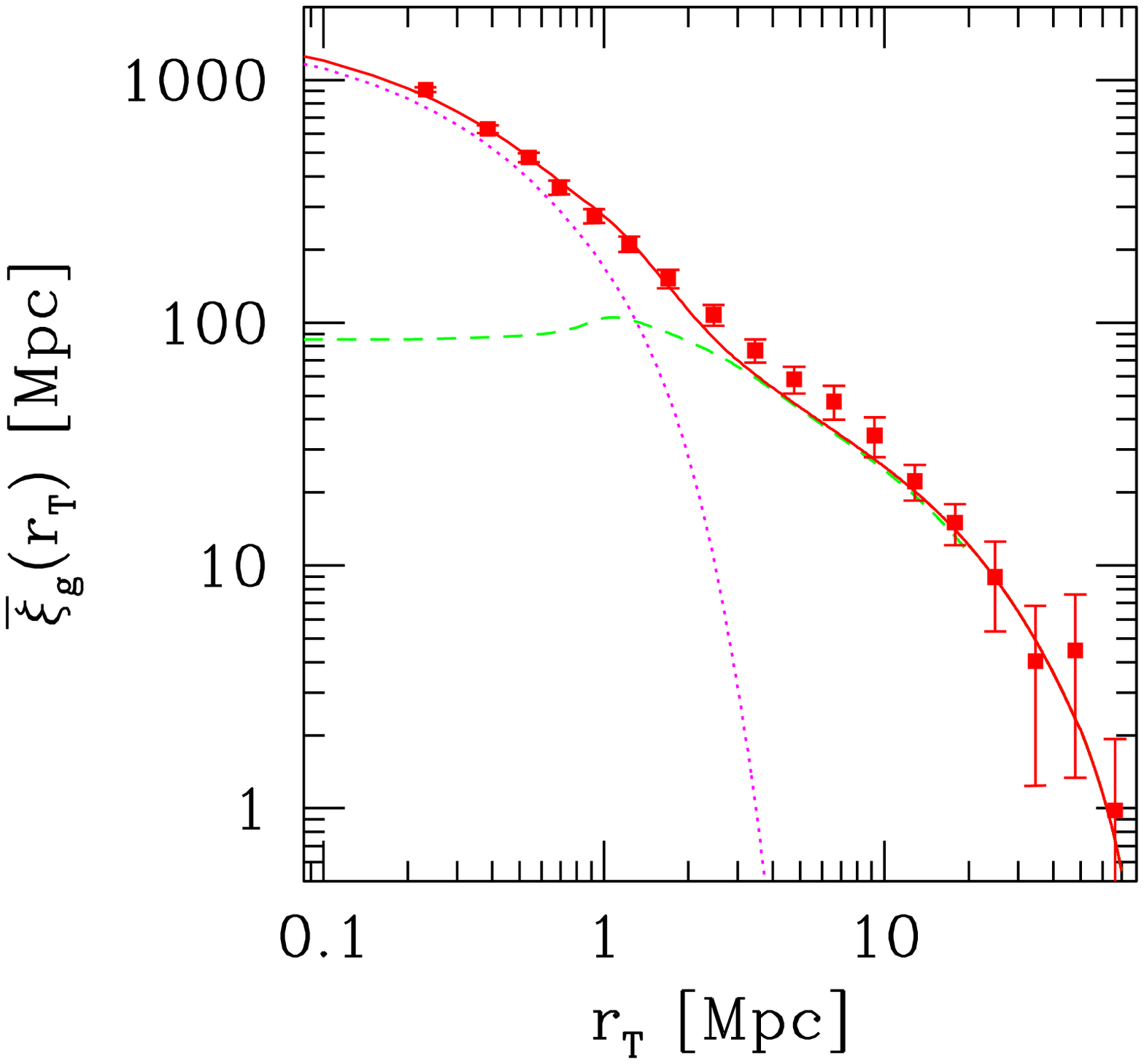}

\includegraphics{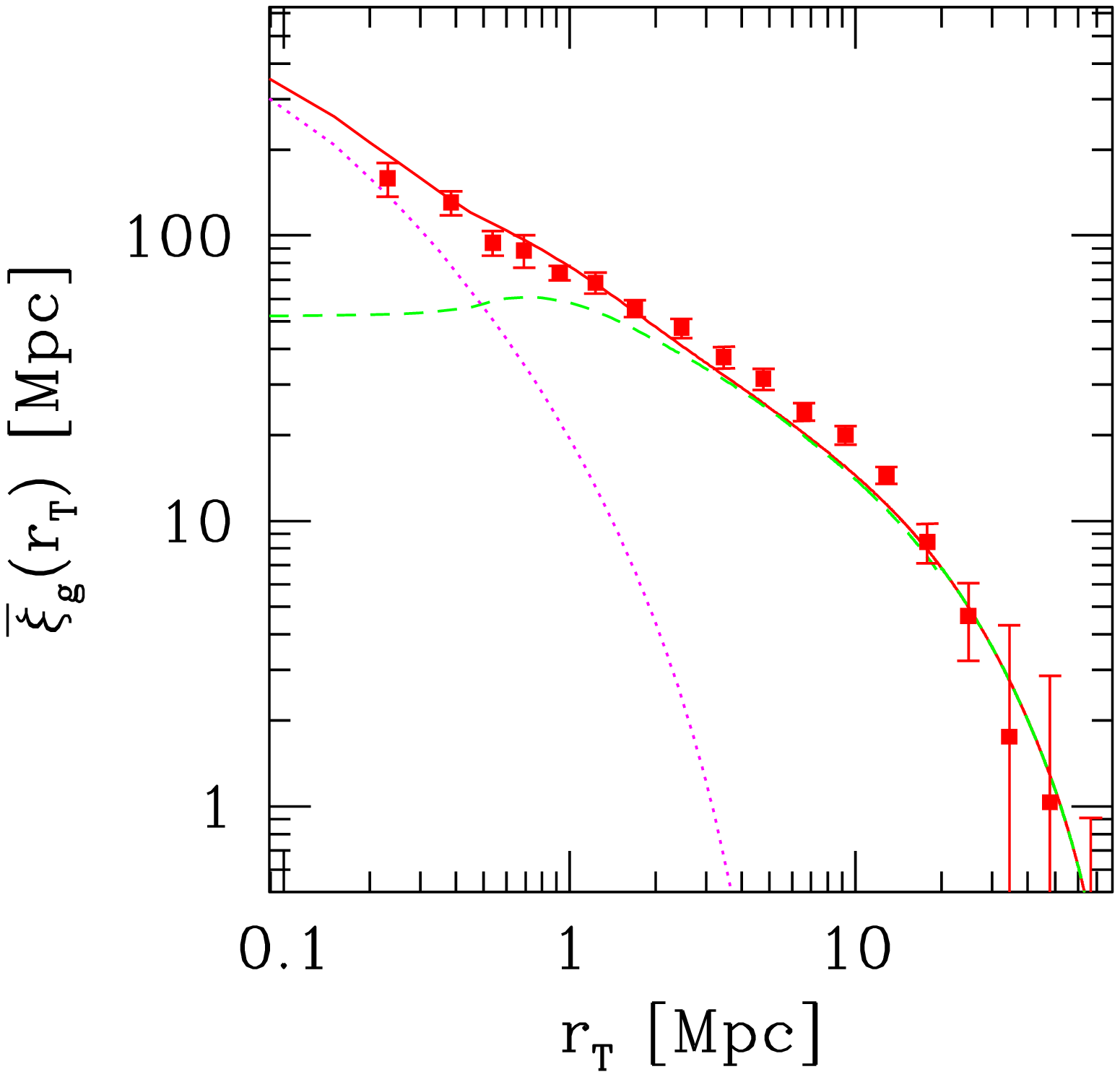}

\sidebyside
{\caption{Projected correlation function of early-type galaxies. Data-points 
represent the results from Madgwick et al. (2003), while the solid curve 
is the best fit to the measurements obtained for a halo number density of the 
form broken power-law, with $\alpha_1=-0.2$, $\alpha_2=1.1$, $m_{\rm cut}=10^{12.6}
m_\odot$,
$m_0=10^{13.5}m_\odot$ and for galaxies distributed within their dark matter 
haloes according to a NFW profile. Dashed 
and dotted lines respectively indicate 
the contribution $\xi_g^{2h}$ from galaxies residing in different haloes and 
the $\xi_g^{1h}$ term originating from galaxies within the same halo.}}
{\caption{Projected correlation function of late-type galaxies. Data-points 
represent the results from Madgwick et al. (2003), while the solid curve 
is the best fit to the measurements obtained for a halo number density of the 
form broken power-law, with $\alpha_1=-0.4$, $\alpha_2=0.7$, $m_{\rm cut}=10^{11}m_\odot$,
$m_0=10^{11.4}m_\odot$ and for galaxies distributed within their dark matter 
haloes according to a NFW profile with $r_{\rm cut}=2\cdot r_{\rm vir}$. 
Dashed and dotted lines respectively indicate 
the contribution $\xi_g^{2h}$ from galaxies residing in different haloes and 
the $\xi_g^{1h}$ term originating from galaxies within the same halo.}}
\end{figure}

The main conclusions of this work are as follows:
 \begin{enumerate}
\item Early-type galaxies (see Figure 1) are well described by a halo occupation 
number of the form broken power-law with $\alpha_1\simeq -0.2$, $\alpha_2\simeq 1.1$, 
$m_{\rm cut}\simeq 10^{12.6}m_\odot$ and $m_0\simeq 10^{13.5}m_\odot$, where the two 
quantities which determine the intermediate-to-high mass behaviour of 
$\langle N_{\rm gal}\rangle$ are measured with a good accuracy.
\item No model can provide a reasonable fit to the correlation function 
of late-type galaxies since they all show an excess of power with respect 
to the data on scales $0.5\simlt r/[\rm Mpc]\simlt 2$. In order to obtain 
an acceptable description of the observations, one has to assume that 
star-forming galaxies are distributed within haloes of masses comparable 
to those of groups and clusters up to two virial radii. This result  
is consistent with the phenomenon of morphological segregation whereby 
late-type galaxies are mostly found in the outer regions of groups or 
clusters (extending well beyond their virial 
radii),  while passive objects preferentially sink into their centres.
\item With the above result in mind, one finds that late-type galaxies 
(see Figure 2) can be described by a halo occupation number of the 
form single power-law with $\alpha_2\simeq 0.7$, 
$m_{\rm cut}\simeq 10^{11}m_\odot$ and $m_0\simeq 10^{11.4}m_\odot$, where the 
quantities which describe $\langle N_{\rm gal}\rangle$ in the high-mass regime 
are determined with a high degree of accuracy.
\item Within the framework of our models, galaxies of any kind seem to follow the 
underlying distribution of dark matter
within haloes as they present the same degree of spatial concentration.
In fact the data indicates both early-type and late-type galaxies to be 
distributed 
within their host haloes according to NFW profiles.
We note however that, even though early-type galaxies can also be described by
means of a shallower distribution of the form $\rho(r)\propto
r^{-\beta}$ with $\beta=2$, this cannot be accepted as a fair modelling of the data 
in the case
of late-type galaxies which instead allow for somehow steeper ($\beta\simeq 2.5)$
profiles. In no case a $\beta=3$ density run can provide an acceptable description 
of the observed correlation function.
These conclusions depend somehow on assuming a specific functional form
for the second moment of the halo occupation distribution. However, Magliocchetti 
\& Porciani (2003) have shown that there is not much freedom in the choice of this 
function if one wants to accurately match the observational data. 
\end{enumerate}

An interesting point to note is that results on the spatial distribution 
of galaxies within haloes and on their halo occupation number are 
independent from each other. There is no degeneracy in the determination of 
$\langle N_{\rm gal}\rangle$ and $\rho(r)$ as they dominate the behaviour 
of the two-point correlation function $\xi_g$ at different scales. Different 
distribution profiles in fact principally determine the slope of 
$\xi_g$ on small enough ($r\simlt 1$~Mpc) scales which probe the inner 
regions of the haloes, while the halo occupation number is mainly responsible 
for the overall normalization of $\xi_g$ and for its slope on large-to-intermediate 
scales. 

Our analysis shows that late-type galaxies can be hosted in haloes with masses 
smaller than it is the case for early-type objects. This is probably due 
to the fact that early-type galaxies are on average more massive (where 
the term here refers to stellar mass) than star-forming objects, especially 
if one considers the population of irregulars, and points to a relationship  
between stellar mass of galaxies and mass of the dark matter haloes which host 
them.

\end{document}